\documentclass[fleqn,usenatbib]{mnras}

\usepackage{newtxtext,newtxmath}

\usepackage[T1]{fontenc}

\DeclareRobustCommand{\VAN}[3]{#2}
\let\VANthebibliography\thebibliography
\def\thebibliography{\DeclareRobustCommand{\VAN}[3]{##3}\VANthebibliography}

\usepackage{graphicx}	
\usepackage{amsmath}	

\newcommand{\ccndopplereqn}{\omega_{\rm obs} \approx \gamma_{\rm em}^{2} \omega_{\rm ns} \label{eq:ccn_doppler_eqn}}
\newcommand{\ccnemittinglorentzfactor}{\gamma_{\rm em} \approx 5.8 \cdot 10^{3} \tilde{\omega}_{\rm ns}^{-0.5} \tilde{\omega}_{\rm obs}^{0.5} \label{eq:ccn_emitting_lorentz_factor}}
\newcommand{\ccnrelativisticduration}{t_{\rm obs} \approx \frac{r_{\rm em}}{\gamma_{\rm em}^{2} c} \label{eq:ccn_relativistic_duration}}
\newcommand{\ccnemittingradius}{r_{\rm em} \approx 1.0 \cdot 10^{15} \tilde{\omega}_{\rm obs} \tilde{t}_{\rm obs} \tilde{\omega}_{\rm ns}^{-1} \, \rm cm \label{eq:ccn_emitting_radius}}
\newcommand{\winddensity}{n \approx \frac{\dot{N}}{c r^{2}} \label{eq:wind_density}}
\newcommand{\pprevanescentthreshold}{\dot{N} > \dot{N}_{\rm ev} \approx 3.6 \cdot 10^{42} \tilde{\omega}_{\rm ns}^{-1} \tilde{\omega}_{\rm obs}^{3} \tilde{t}_{\rm obs}^{2} \, \rm s^{-1} \label{eq:ppr_evanescent_threshold}}
\newcommand{\ccnfluence}{S \approx \frac{\gamma_{\rm em}^{2} b_{\rm em}^{2} r_{\rm em}^{3}}{\omega_{\rm obs} d^{2}} \label{eq:ccn_fluence}}
\newcommand{\ccnemittingmagneticfieldevanescent}{B_{\rm em} \approx 1.6 \cdot 10^{-7} \tilde{d} \tilde{S}^{0.5} \tilde{\omega}_{\rm ns}^{2} \tilde{\omega}_{\rm obs}^{-1.5} \tilde{t}_{\rm obs}^{-1.5} \, \rm G  \label{eq:ccn_emitting_magnetic_field_evanescent}}
\newcommand{\ccnemittingmagneticfieldperdurable}{B_{\rm em} \approx 1.6 \cdot 10^{-7} \tilde{\omega}_{\rm ns} \tilde{d} \tilde{S}^{0.5} \tilde{\omega}_{\rm obs}^{1.5} \tilde{t}_{\rm obs}^{0.5} \frac{3.6 \cdot 10^{42} \, {\rm s}^{-1}}{\dot{N}}  \, \rm G  \label{eq:ccn_emitting_magnetic_field_perdurable}}
\newcommand{\nsinnermagneticfield}{B_{\rm in} \approx \frac{B_{\rm ns} R_{\rm ns}^{3}}{r^{3}} \label{eq:ns_inner_magnetic_field}}
\newcommand{\nsoutermagneticfield}{B_{\rm out} \approx \frac{B_{\rm ns} R_{\rm ns}^{3} \omega_{\rm ns}^{2}}{c^{2} r} \label{eq:ns_outer_magnetic_field}}
\newcommand{\ccnneutronstarmagneticfieldevanescent}{B_{\rm ns} \approx 1.6 \cdot 10^{8} \tilde{d} \tilde{S}^{0.5} \tilde{\omega}_{\rm ns}^{-1} \tilde{\omega}_{\rm obs}^{-0.5} \tilde{t}_{\rm obs}^{-0.5} \, \rm G \label{eq:ccn_neutron_star_magnetic_field_evanescent}}
\newcommand{\ccnneutronstarmagneticfieldperdurable}{B_{\rm ns} \approx 1.6 \cdot 10^{8} \tilde{d} \tilde{S}^{0.5} \tilde{\omega}_{\rm ns}^{-2} \tilde{\omega}_{\rm obs}^{2.5} \tilde{t}_{\rm obs}^{1.5} \frac{3.6 \cdot 10^{42} \, {\rm s}^{-1}}{\dot{N}}  \, \rm G \label{eq:ccn_neutron_star_magnetic_field_perdurable}}
\newcommand{\ccnpprthresholdformagneticdominance}{\dot{N} < \dot{N}_{\rm md} \approx 1.0 \cdot 10^{33} \tilde{S} \tilde{\omega}_{\rm ns}^{2} \tilde{\omega}_{\rm obs}^{-1} \tilde{d}^{2} \tilde{t}_{\rm obs}^{-1} {\rm s^{-1}} \label{eq:ccn_ppr_threshold_for_magnetic_dominance}}
\newcommand{\ccnmagneticdominatedenergy}{E \approx 9.0 \cdot 10^{38} \tilde{S} \tilde{\omega}_{\rm obs} \tilde{d}^{2} \, \rm erg \label{eq:ccn_magnetic_dominated_energy}}
\newcommand{\ccnmatterdominatedenergy}{E \approx 9.0 \cdot 10^{38} \tilde{t}_{\rm obs} \tilde{\omega}_{\rm ns}^{-2} \tilde{\omega}_{\rm obs}^{2} \frac{\dot{N}}{1.0 \cdot 10^{33} \, {\rm s}^{-1}} \, \rm erg \label{eq:ccn_matter_dominated_energy}}
\newcommand{\ccnlightcylinderfrequency}{\omega_{\rm lc} \approx 1.0 \cdot 10^{15} \tilde{t}_{\rm obs} \tilde{\omega}_{\rm obs}^{2} \, \rm Hz \label{eq:ccn_light_cylinder_frequency}}
\newcommand{\ccnthomsonopticaldepth}{\tau_{\rm t} \approx 2.6 \cdot 10^{-18} \tilde{\omega}_{\rm ns} \tilde{\omega}_{\rm obs}^{-1} \tilde{t}_{\rm obs}^{-1}\frac{\dot{N}}{1.0 \cdot 10^{33} \, {\rm s}^{-1}} \label{eq:ccn_thomson_optical_depth}}
\newcommand{\ccninversecomptonfrequency}{\omega_{\rm IC} \approx 3.3 \cdot 10^{16} \tilde{\omega}_{\rm ns}^{-1} \tilde{\omega}_{\rm obs}^{2} \, \rm Hz \label{eq:ccn_inverse_compton_frequency}}
\newcommand{\ccninversecomptonflux}{f_{\rm IC} \approx 1.5 \cdot 10^{-28} \tilde{S} \tilde{\omega}_{\rm obs} \tilde{t}_{\rm obs}^{-1} \Delta t_{\rm UV}^{-1}\frac{\dot{N}}{1.0 \cdot 10^{33} \, {\rm s}^{-1}} \rm \frac{erg}{cm^2 g} \label{eq:ccn_inverse_compton_flux}}
\newcommand{\ccninversecomptonmagnitude}{57 \label{eq:ccn_inverse_compton_magnitude}}
\newcommand{\ccnsynchrotronobservedfrequency}{\omega_{\rm syn, obs} \approx 5.6 \cdot 10^{11} \tilde{d} \tilde{S}^{0.5} \tilde{\omega}_{\rm ns}^{0.5} \tilde{t}_{\rm obs}^{-1.5} \, \rm Hz \label{eq:ccn_synchrotron_observed_frequency}}
\newcommand{\ccnsynchrotronobservedspectralflux}{f_{\rm syn} \approx 5.7 \cdot 10^{-21} \tilde{S}^{0.3} \tilde{\omega}_{\rm ns}^{0.3} \tilde{\omega}_{\rm obs}^{0.3} \tilde{d}^{-1.3}\frac{\dot{N}}{1.0 \cdot 10^{33} \, {\rm s}^{-1}} \, \rm Jy \label{eq:ccn_synchrotron_observed_spectral_flux}}
\newcommand{\cnicdopplerequation}{\omega_{\rm obs} \approx \gamma_{\rm em}^{4} \omega_{\rm ns} \label{eq:cnic_doppler_equation}}
\newcommand{\cnicemittinglorentzfactor}{\gamma_{\rm em} \approx 7.6 \cdot 10^{1} \tilde{\omega}_{\rm ns}^{-0.25} \tilde{\omega}_{\rm obs}^{0.25} \label{eq:cnic_emitting_lorentz_factor}}
\newcommand{\cnicemittingradius}{r_{\rm em} \approx 1.7 \cdot 10^{11} \tilde{t}_{\rm obs} \tilde{\omega}_{\rm ns}^{-0.5} \tilde{\omega}_{\rm obs}^{0.5} \, \rm cm \label{eq:cnic_emitting_radius}}
\newcommand{\cnicshockedframecnfrequency}{\tilde{\omega}_{\rm CN}' \approx 2.3 \cdot 10^{3} \tilde{\omega}_{\rm ns}^{0.75} \tilde{\omega}_{\rm obs}^{0.25} \, \rm Hz \label{eq:cnic_shocked_frame_cn_frequency}}
\newcommand{\cnicpprevanescentthreshold}{\dot{N}_{\rm ev} \approx 1.9 \cdot 10^{31} \tilde{\omega}_{\rm ns}^{0.5} \tilde{\omega}_{\rm obs}^{1.5} \tilde{t}_{\rm obs}^{2} \, \rm  s^{-1} \label{eq:cnic_ppr_evanescent_threshold}}
\newcommand{\cniccnenergyperdurable}{E_{\rm CN,pe} \approx \frac{B_{\rm em}^{2} \dot{N}^{2} c r_{e}^{2}}{\gamma_{\rm em}^{2} \omega_{\rm ns}^{\frac{7}{2}} \sqrt{\omega_{\rm obs}} t_{\rm obs}} \label{eq:cnic_cn_energy_perdurable}}
\newcommand{\cnicenergyperdurable}{E_{\rm pe} \approx \frac{B_{\rm em}^{2} \dot{N}^{3} r_{e}^{4}}{\omega_{\rm ns}^{3} \omega_{\rm obs} c t_{\rm obs}^{2}} \label{eq:cnic_energy_perdurable}}
\newcommand{\cnicfluenceperdurable}{S_{\rm pe} \approx \frac{B_{\rm em}^{2} \dot{N}^{3} r_{e}^{4}}{\omega_{\rm ns}^{3} \omega_{\rm obs}^{2} c d^{2} t_{\rm obs}^{2}} \label{eq:cnic_fluence_perdurable}}
\newcommand{\cnicemittingmagneticfieldperdurable}{B_{\rm em, ev} \approx 5.7 \cdot 10^{7} \tilde{d} \tilde{S}^{0.5} \tilde{\omega}_{\rm ns}^{1.125} \tilde{\omega}_{\rm obs}^{-0.625} \tilde{t}_{\rm obs}^{-1}\left(\frac{\dot{N}}{1.9 \cdot 10^{31} {\rm s}^{-1}}\right)^{-0.25} \, \rm G \label{eq:cnic_emitting_magnetic_field_perdurable}}
\newcommand{\cnicneutronstarmagneticfieldperdurable}{B_{\rm ns, pe} \approx 7.5 \cdot 10^{17} \tilde{d} \tilde{S}^{0.5} \tilde{\omega}_{\rm ns}^{-1} \tilde{\omega}_{\rm obs}^{1.5} \tilde{t}_{\rm obs}^{2}\left(\frac{\dot{N}}{1.9 \cdot 10^{31} {\rm s}^{-1}}\right)^{-1.5} \, \rm G \label{eq:cnic_neutron_star_magnetic_field_perdurable}}
\newcommand{\cniccnenergyevanescent}{E_{\rm CN,ev} \approx \frac{B_{\rm em}^{2} \omega_{\rm obs}^{2} c^{3} t_{\rm obs}^{3}}{\omega_{\rm ns}^{2}} \label{eq:cnic_cn_energy_evanescent}}
\newcommand{\cnicenergyevanescent}{E_{\rm ev} \approx \frac{B_{\rm em}^{2} \sqrt{\dot{N}} \omega_{\rm obs}^{\frac{9}{4}} c^{\frac{3}{2}} r_{e}^{\frac{3}{2}} t_{\rm obs}^{2}}{\omega_{\rm ns}^{\frac{9}{4}}} \label{eq:cnic_energy_evanescent}}
\newcommand{\cnicfluenceevanescent}{S_{\rm ev} \approx \frac{B_{\rm em}^{2} \sqrt{\dot{N}} \omega_{\rm obs}^{\frac{5}{4}} c^{\frac{3}{2}} r_{e}^{\frac{3}{2}} t_{\rm obs}^{2}}{\omega_{\rm ns}^{\frac{9}{4}} d^{2}} \label{eq:cnic_fluence_evanescent}}
\newcommand{\cnicneutronstarmagneticfieldevanescent}{B_{\rm ns, ev} \approx 9.8 \cdot 10^{18} \tilde{d} \tilde{S}^{0.5} \tilde{\omega}_{\rm ns}^{-1.375} \tilde{\omega}_{\rm obs}^{-0.125}\left(\frac{\dot{N}}{1.9 \cdot 10^{31} {\rm s}^{-1}}\right)^{-0.25} \, \rm G \label{eq:cnic_neutron_star_magnetic_field_evanescent}}
\newcommand{\cnicpprmagneticdominanceperdurable}{\dot{N}_{\rm md, pe} \approx 6.2 \cdot 10^{41} \tilde{d} \tilde{S}^{0.5} \tilde{\omega}_{\rm ns}^{0.5} \, \rm s^{-1} \label{eq:cnic_ppr_magnetic_dominance_perdurable}}
\newcommand{\cnicpprmagneticdominanceevanescent}{\dot{N}_{\rm md, ev} \approx 6.2 \cdot 10^{46} \tilde{S}^{0.7} \tilde{\omega}_{\rm ns}^{0.83} \tilde{\omega}_{\rm obs}^{-0.17} \tilde{d}^{1.3} \label{eq:cnic_ppr_magnetic_dominance_evanescent}}
\newcommand{\cnicexplicitcnenergyperdurable}{E_{\rm CN, pe} \approx 5.6 \cdot 10^{50} \tilde{S} \tilde{\omega}_{\rm obs} \tilde{t}_{\rm obs} \tilde{d}^{2}\left(\frac{\dot{N}}{1.9 \cdot 10^{31} {\rm s}^{-1}}\right)^{- 1} \, \rm erg \label{eq:cnic_explicit_cn_energy_perdurable}}
\newcommand{\cnicexplicitcnenergyevanescent}{E_{\rm CN, ev} \approx 9.6 \cdot 10^{52} \tilde{S} \tilde{t}_{\rm obs} \tilde{\omega}_{\rm ns}^{0.25} \tilde{\omega}_{\rm obs}^{0.75} \tilde{d}^{2}\left(\frac{\dot{N}}{1.9 \cdot 10^{31} {\rm s}^{-1}}\right)^{- 0.5} \, \rm erg \label{eq:cnic_explicit_cn_energy_evanescent}}
\newcommand{\cniclightcylinderfrequency}{\omega_{\rm lc} \approx 3.0 \cdot 10^{13} \tilde{\omega}_{\rm ns} \tilde{\omega}_{\rm obs}^{2} \tilde{t}_{\rm obs}^{2} \, \rm Hz \label{eq:cnic_light_cylinder_frequency}}
\newcommand{\cnicdoublecomptonfrequency}{\omega_{\rm 2IC} \approx 5.8 \cdot 10^{12} \tilde{\omega}_{\rm ns}^{-0.5} \tilde{\omega}_{\rm obs}^{1.5} \, \rm Hz \label{eq:cnic_double_compton_frequency}}
\newcommand{\cnicshockframesynchrotroncoolingtime}{t'_{\rm syn} \approx 2.4 \cdot 10^{-13} \tilde{S}^{-1} \tilde{\omega}_{\rm ns}^{-1.5} \tilde{\omega}_{\rm obs}^{0.5} \tilde{d}^{-2} \tilde{t}_{\rm obs}^{2}\left(\frac{\dot{N}}{1.9 \cdot 10^{31} {\rm s}^{-1}}\right)^{0.5} \, \rm s \label{eq:cnic_shock_frame_synchrotron_cooling_time}}
\newcommand{\cnicsynchrotronfastcoolingenergy}{E_{\rm fc} \approx 8.7 \cdot 10^{25} \tilde{t}_{\rm obs} \tilde{\omega}_{\rm ns}^{-0.5} \tilde{\omega}_{\rm obs}^{0.5}\frac{\dot{N}}{1.9 \cdot 10^{31} {\rm s}^{-1}} \, \rm erg \label{eq:cnic_synchrotron_fast_cooling_energy}}
\newcommand{\cnicobservedsynchrotronfrequency}{\omega_{\rm syn, obs} \approx 4.4 \cdot 10^{20} \tilde{d} \tilde{S}^{0.5} \tilde{\omega}_{\rm ns}^{0.375} \tilde{\omega}_{\rm obs}^{0.125} \tilde{t}_{\rm obs}^{-1}\left(\frac{\dot{N}}{1.9 \cdot 10^{31} {\rm s}^{-1}}\right)^{-0.25} \, \rm Hz \label{eq:cnic_observed_synchrotron_frequency}}
\newcommand{\cnicobservedcyclotronfrequency}{\omega_{\rm cyc, obs} \approx 7.6 \cdot 10^{16} \tilde{d} \tilde{S}^{0.5} \tilde{\omega}_{\rm ns}^{0.875} \tilde{\omega}_{\rm obs}^{-0.375} \tilde{t}_{\rm obs}^{-1}\left(\frac{\dot{N}}{1.9 \cdot 10^{31} {\rm s}^{-1}}\right)^{-0.25} \, \rm Hz \label{eq:cnic_observed_cyclotron_frequency}}

\title[electromagnetic mirror model for FRBs]{The Moving Mirror model for Fast Radio Bursts}

\author[A. Yalinewich and U-L. Pen]{
Almog Yalinewich,$^{1}$\thanks{E-mail: almog.yalin@gmail.com   }
U-L. Pen,$^{1}$
\\
$^{1}$Canadian Institute for Theoretical Astrophysics, 60 St. George St., Toronto, ON M5S 3H8, Canada 
}

\date{Accepted XXX. Received YYY; in original form ZZZ}

\pubyear{2022}

\begin{document}
\label{firstpage}
\pagerange{\pageref{firstpage}--\pageref{lastpage}}
\maketitle

\begin{abstract}

Recent observations of coherent radiation from the Crab pulsar (Bij
et al 2021) suggest the emission is driven by an ultra - relativistic ($\gamma \sim 10^4$),
cold plasma flow. A relativistically expanding plasma shell can compress the ambient magnetic field, like a moving mirror, and thus produce coherent radiation whose wavelength is shorter than that of the ambient medium by $\gamma^2$. This mechanism has been studied in the past by Colgate and Noerdelinger (1971), in the context of radio loud supernova explosions. In this work we propose that a similar mechanism drives the coherent emission in fast radio bursts. The high Lorenz factors dramatically
lower the implied energy and magnetic field requirements, allowing the spin down energy
of regular (or even recycled), fast spinning pulsars, rather than slow spinning magnetars, to explain FRBs. We show that this model can explain the frequency and the time
evolution of observed FRBs, as well as their duration, energetics
and absence of panchromatic counterparts. 
We also predict that the peak frequency of sub pulses decline with
observation time as $\omega_{\rm obs} \propto t_{\rm obs}^{-1/2}$.
Unfortunately, with current capabilities it is not possible to constrain the shape of the curve $\omega_{\rm obs} \left(t_{\rm obs} \right)$. Finally, we find that a variation of this model can explain weaker radio transients, such as the one observed from a galactic magnetar. In this variant, the shock wave produces low frequency photons which are then Compton scattered to the GHz range.

\end{abstract}

\begin{keywords}
keyword1 -- keyword2 -- keyword3
\end{keywords}



\section{Introduction}

Fast radio bursts, as the name implies, are short flashes of radio
emission from outer space. These radio transients produce coherent radio emission, as evidenced by the observed strong
scintillation \citep{Masui2015DenseBurst}, implying coherent wave
optics lensing \citep{2018NatAs...2..842P}.  This potentially enables
the precision measurement of intervening stars,
planets \citep{2020MNRAS.497.4956J}, and potentially dark matter and
dark energy \citep{2021A&A...645A..44W}. The inferred high instantaneous
luminosity has led to a wild speculation on the nature of the
central engine \citep[see][and references therein]{Platts2018ABursts}.
Taken at face value, the inferred brightness temperatures exceed the
Planck Temperature, and local electric field suggests a local
breakdown of the QED vacuum through the Schwinger mechanism
\citep{2015arXiv150506400S}.  Existing models of coherent emission
involve complex non-ideal plasma effects, which depend on unknown
environmental properties, and uncertain theoretical inferences.

Recently, a number of observations hinted that FRBs are produced by
neutron stars. First, a bright radio pulse was detected from a
galactic magnetar \citep{Andersen2020AMagnetar,2020Natur.587...59B}. The frequency and
duration of this burst was typical for FRBs, but the flux was
considerably lower.  Second, giant pulses were detected from the crab
pulsar which have the comparable duration and frequency as FRBs, but
lower flux \citep{Bij2021KinematicsPulses}. The latter also indicated
that the giant pulses originate on the light cylinder, and involve a
relativistic motion with a coherent Lorentz factor of around
$\gamma \sim 10^4$.  The high Lorenz factor enables relativistic
beaming, reducing the inferred instantaneous luminosities by up to
$\gamma^2$. 

Prompted by these observations, in this work we consider a new model
for FRB emission. In our model, a sudden release of energy occurs close to
the light cylinder, possibly due to magnetic reconnection
\citep{Philippov2019PulsarReconnection}. As a result, a plasma shell
is launched at a relativistic velocity. This relativistic shell
compresses the ambient dipolar field of the neutron star, and thus
emit coherent radio emission. This model is based on a previous
work that explored the same mechanism in the context of radio
emission from supernova explosions \citep{Colgate1971CoherentShells}.

The plan of this paper is as follows. In section \ref{sec:ccn} we describe our model, and calibrate its parameters according to the observational data. In section \ref{sec:icnc} we consider a variant of the model, in which the shock wave emits coherent radiation at a low frequency, which are then Compton scattered to the GHz range. Finally, in section \ref{sec:conclusions} we discuss the results.

\section{Classical Colgate Noerdelinger Mechanism} \label{sec:ccn}

\subsection{Main Emission}

In this section we explore whether fast radio bursts can be produced by the mechanism described by \cite{Colgate1971CoherentShells}. In their model, a relativistic plasma shell compresses a pre - existing magnetic field and thus produces coherent radio emission. In our case the magnetic field originates from a misaligned neutron star (i.e. the magnetic dipole is misaligned with respect to the rotation axis) rotating with a frequency $\omega_{\rm ns}$. This field is carried outward by a striped wind at some fraction of the speed of light, and so the magnetic field oscillates with a wavelength comparable to the radius of the light cylinder $c/\omega_{\rm ns}$ \citep{Mochol2017PulsarWinds}. When the shock wave encounters the oscillating magnetic field, it produces a radiation with a frequency
\begin{equation}
  \ccndopplereqn
\end{equation}
where $\gamma_{\rm em}$ is the Lorentz factor of the shock wave at the emitting radius. An observer will detect a signal with a characteristic frequency $\omega_{\rm ns}$. Using typical values for the observed frequency and neutron star rotation frequency we can estimate the Lorentz factor
\begin{equation}
  \ccnemittinglorentzfactor
\end{equation}
where $\tilde{\omega}_{\rm ns} = \omega_{\rm ns} / 30 \, \rm Hz$ and $\tilde{\omega}_{\rm obs} = \omega_{\rm obs} / 1 \, \rm GHz$. We chose our fiducial neutron star rotation frequency to be the same as the Crab pulsar \citep{Lyne199323History.}.

If the emission originates from a sphere of radius $r$ around the neutron star, then the de-dispersed observed duration of the singal will be smaller than the light crossing time by a factor of $\gamma^2$
\begin{equation}
  \ccnrelativisticduration
\end{equation}
Solving for the emitting radius $r_{\rm em}$ yields
\begin{equation}
  \ccnemittingradius
\end{equation}
where $\tilde{t}_{\rm obs} = t_{\rm obs} / 1 \, \rm ms$.

We can also use the observed fluence to calibrate the parameters of the model, though this relation will be more complicated than the previous ones. In the original work by \citep{Colgate1971CoherentShells}, the authors assumed that all the ambient magnetic field is swept up by the shock wave. In order to get a complete reflection, the plasma frequency of the shocked plasma must exceed the frequency of radiation in the rest frame. In this case the wave transmitted into shocked plasma shell will be evanescent, and so we call this evanescent regime. The plasma frequency is given by $q \sqrt{n/m_e} \approx c \sqrt{n r_e}$ where $q$ is the elementary charge. The downstream rest density is larger by a factor of $\gamma$ relative to the upstream, but the interia of the electrons also increases by the same factor, so that the plasma frequency downstream is the same as the upstream. In the downstream frame the frequency of radiation is $\omega_{\rm obs} / \gamma$. The condition for being in the evanescent regime
\begin{equation}
    c \sqrt{n r_e} > \gamma \omega_{\rm ns} \, . \label{eq:evanescent_condition}
\end{equation}
If condition \ref{eq:evanescent_condition} is not satisfied, then only a small fraction of the ambient magnetic field is converted to radiation. The refractive index of plasma is given by $\sqrt{1 - \omega_{\rm pl}^2/\omega^2} \approx 1 - \omega_{\rm pl}^2/ 2 \omega^2$, so the reflection coefficient is given by
\begin{equation}
    \mathcal{R} \approx \frac{\omega_{\rm pl}'^2}{\omega_{\rm obs}'^2} \approx \frac{c^2 n r_e}{\gamma^2 \omega_{\rm ns}^2}
\end{equation}
where a prime indicates that the quantity pertains to the shell's rest frame. The energy therefore decreases by a factor of $\mathcal{R}^2$ relative to the evanescent case. 

We denote the rate at which the neutron star emits particles (primarily electrons and positrons) by $\dot{N}$. The density at a distance $r$ is given by
\begin{equation}
  \winddensity 
\end{equation}
Using relations \ref{eq:wind_density} and \ref{eq:evanescent_condition} we can obtain the minimal $\dot{N}$ for the evanescent regime
\begin{equation}
  \pprevanescentthreshold
\end{equation}

The fluence $S$ observed at a distance $d$ is given by
\begin{equation}
  \ccnfluence
\end{equation}
where
\begin{equation}
    b_{\rm em} \approx B_{\rm em} \cdot \min \left(\frac{\dot{N}}{\dot{N}_{\rm ev}},1\right)
\end{equation}
and $B_{\rm em}$ is the magnetic field in the emitting region. In the evanescent regime $\dot{N} > \dot{N}_{\rm ev}$ the magnetic field in the emitting region is given by
\begin{equation}
  b_{\rm em} \approx \ccnemittingmagneticfieldevanescent
\end{equation}
and in the perdurable regime $\dot{N} < \dot{N}_{\rm ev}$ the magnetic field is given by
\begin{equation}
  \ccnemittingmagneticfieldperdurable
\end{equation}
where $\tilde{d} = d / 1 \, \rm Gpc$ and $\tilde{S} = S / 10 \, \rm Jy \, ms$. We chose typical typical values for FRB pulses \citep{Niu2021CRAFTSFAST, Macquart2020ABursts}.

A naive model for the magnetic field \citep{Goldreich1969PulsarElectrodynamics} predicts that dipolar inside the light cylinder
\begin{equation}
  \nsinnermagneticfield
\end{equation}
where $R_{\rm ns}$ is the neutron star's radius and $B_{\rm ns}$ is the magnetic field on its surface. At distances larger than the light cylinder $r > R_{\rm lc} \approx \omega_{\rm ns}/c$ where $\omega_{\rm ns}$ is the neutron star's rotation frequency, the magnetic field is dominated by the toroidal component, which scales as $1/r$, and hence given by
\begin{equation}
  \nsoutermagneticfield \, .
\end{equation}
Using this expression, we can estimate the magnetic field on the surface of the neutron star. In the evanescent regime, the magnetic field is given by
\begin{equation}
  \ccnneutronstarmagneticfieldevanescent
\end{equation}
and in the perdurable regime
\begin{equation}
  \ccnneutronstarmagneticfieldperdurable
\end{equation}
where $\tilde{R}_{\rm ns} = R_{\rm ns} / 10 \, \rm km$. The minimum value of the magnetic field is typical of recycled millisecond pulsars \citep{Cruces2019OnAccretion}.

The amount of energy required to drive such a shock wave depends on whether the energy density of the magnetic field exceeds the rest mass energy of the wind. The condition for magnetic dominance is given by $b_{\rm em}^2 > n_{\rm em} m_e c^2$. Solving for the particle production rate yields
\begin{equation}
\ccnpprthresholdformagneticdominance \, .
\end{equation}
Since this particle emission rate is lower than the critical value for evanescent in equation \ref{eq:ppr_evanescent_threshold}, magnetic dominance can only occur in the perdurable regime.

The shock trajectory in this environment is given by the Blandford McKee solution \citep{Blandford1976FluidWaves}
\begin{equation}
    \gamma \approx \sqrt{E / \max\left(b_{\rm em}^2 r_{\rm em}^3, m_e c \dot{N} r_{\rm em} \right)} \label{eq:trajectory}
\end{equation}
where $m_e$ is the electron rest mass. In the magnetic dominated regime $\dot{N} < \dot{N}_{\rm md}$ the energy of the explosion is
\begin{equation}
  \ccnmagneticdominatedenergy \, .
\end{equation}
In the matter dominated regime $\dot{N} > \dot{N}_{\rm md}$ the energy is given by
\begin{equation}
  \ccnmatterdominatedenergy \, .
\end{equation}

As time progresses, the shock wave sweeps across more leptons and radiates more energy, and thus decelerates. Since both the rest mass energy of the wind and the magnetic pressure decrease as $r^{-2}$, the Lorentz factor decreases as $\gamma \propto r^{-1/2}$. From equation \ref{eq:ccn_doppler_eqn} and \ref{eq:ccn_relativistic_duration} we find that the observed frequency should scale with observation time as $\omega_{\rm obs} \propto t_{\rm obs}^{-0.5}$. This prediction is in qualitative agreement with observations that consistently show sub pulses shifting to lower frequencies \citep{Hessels2019FRBStructure}. Unfortunately, these observations are not sensitive enough to place quantitative constraint the manner of the decline in frequency.
 
Finally, we discuss the polarisation of the emitted radiation. The
polarisation is determined by the direction of the ambient magnetic
field. The observed polarisation depends on orientation of the line of
sight relative to the spin axis and the magnetic dipole. In the case
of an aligned rotator the polarisation will be close to unity when the
line of sight lies in the equatorial plane, and the polarisation
decreases as the line of sight moves closer to the poles. When the
line of sight is close to the rotation axis, but the magnetic dipole
is misaligned, it is possible to get circular
polsarisation. Differences between lines of sight and misalignment
between rotation and dipole field can account for the variety of
polarisation fractions measured in observations
\citep{Mckinven2021ACHIME/FRB}. Moreover, even a perfectly linearly
polarised emission can be depolarised by a scattering
screen \citep{2022Sci...375.1266F}, though this process might give rise
to a non negligible circular polarisation
\citep{Beniamini2022FaradayFRBs}.  Evidence for scattering
depolarization has been observed in the galactic centre
magnetar\citep{2018ApJ...852L..12D}.

\subsection{Counterparts}

In this section we explore radiation emitted from our model other than the coherent radio signal. We divide this into two categories. The first is radiation emitted in other wavelengths, which we call panchromatic. The second is incoherent emission due to synchrotron. The former has to date not been detected, while a persistent synchrotron emission has only been detected in one source \citep{Chatterjee2017AHost}.

\subsubsection{Panchromatic}

According to our model, the Lorentz factor changes continuously. Therefore, the peak frequency also changes continuously. The highest frequency occurs near the light cylinder, and is given by
\begin{equation}
  \ccnlightcylinderfrequency \, .
\end{equation}
This means that the most energetic photons produced by our mechanism are in the UV range, but not in the X ray range, in agreement with the non detection of X ray counterparts \citep{Scholz2017Simultaneous121102}.

Our mechanism does produce a burst of emission in the visible range, but due to its short duration, it is hard to observe. The radiant energy is independent of the emitting radius, and is given by equation TBA. This energy is emitted over a period much shorter than any reasonable exposure time, so the effective luminosity is given by the latter. If we adopt a very optimistic exposure time of $\Delta t_v \approx 1 \, \rm s$, then the effective luminosity would be $10^{39} \, \rm erg/s$. At a distance of $d \approx 1 \, \rm Gpc$, such a luminosity translates to an apparent magnitude of 28. This is much dimmer than current upper bounds on visible counterparts, around a magnitude of 19 \citep{Nunez2021ConstrainingBursts}. Interestingly, if an FRB happens in our galactic neighbourhood ($d \approx 1 \, \rm Mpc$), then the apparent magnitude would just be 13, and can therefore be detected by many time domain missions such as ASASSN, ZTF or VRO. We note that a counterpart in the visible range is not guaranteed, since it requires the energy release to occur close to the light cylinder, and this energy release can occur at much larger radii.

Another mechanism for producing higher frequency photons is through inverse Compton scattering. The Thomson optical depth is 
\begin{equation}
  \ccnthomsonopticaldepth \, .
\end{equation}
The frequency of the inverse Compton radiation will be greater than the radio signal frequency by a factor of $\gamma^2$
\begin{equation}
 \ccninversecomptonfrequency \, . 
\end{equation}
We therefore find that a single inverse Compton scattering boosts the radiation from the radio range to the far UV range. The fluence of the the inverse Compton is greater by a factor of $\gamma^2$ from the radio signal, and smaller by a factor of $\tau_t$. While the instantaneous flux can be high, the duration is expected to be much smaller than a telescope's exposure time, so the effective luminosity will be smaller. The effective flux is therefore given by
\begin{equation}
  \ccninversecomptonflux
\end{equation}
where $\Delta \tilde{t}_{\rm UV} = \Delta t_{\rm UV} / 1 \, \rm minute$ and $\Delta t_{\rm UV}$ is the telescope's exposure time. This flux corresponds to apparent magnitude $\ccninversecomptonmagnitude$, which is too dim for any current or planned telescope. 

\subsubsection{Incoherent Emission}

Relativistic electrons moving in the presence of magnetic fields also produce incoherent synchrotron emission. In this section we will estimate the luminosity of this emission. The luminosity of a single electron is given by
\begin{equation}
    L_1 \approx c B'^2 r_e^2
\end{equation}
where $B' = \gamma B$ is the magnetic field in the shell frame. The number of radiating electrons is $N \approx n r^3$, so the total luminosity in the shell's rest frame is $L_1 N$. The boosted luminosity an observer will see is $\gamma^2 L_1 N$. The peak of the synchrotron frequency in the rest frame is given by $\gamma^2 B \sqrt{r_e/m_e}$, and in the observer frame it is bigger by another Lorentz factor, therefore
\begin{equation}
  \ccnsynchrotronobservedfrequency \, .
\end{equation}
Below the peak, the spectral flux declines with frequency as $\omega^{1/3}$ \citep{Rybicki1979RadiativeAstrophysics}, so the spectral flux at the FRB frequency is given by 
\begin{equation}
  \ccnsynchrotronobservedspectralflux \, .
\end{equation}
We get that the incoherent radio emission is negligible, in accordance with most observations. To date, there is only one FRB associated with a persistent synchrotron radio emitter \citep{Chatterjee2017AHost}.

\section{Colgate Noerdelinger Inverse Compton Mechanism} \label{sec:icnc}

\subsection{Main Emission}

Another possibility of producing radio signals in the GHz range is by first producing photons at lower frequencies by the Colgate Noerdelinger mechanism described above, and then having them Compton scattered to higher frequencies. The Colgate Noerdelinger mechanism boosts the frequency by a factor of $\gamma^2$, and Compton scattering boosts it again by the same factor, such that the relation between the neutron star and observed radiation is given by
\begin{equation}
  \cnicdopplerequation \, .
\end{equation}
Solving for the Lorentz factor at the emitting region $\gamma_{\rm em}$ yields
\begin{equation}
  \cnicemittinglorentzfactor \, .
\end{equation}
We find that the Lorentz factor in the emitting region is considerably smaller than in the previous section. The emitting radius is also smaller
\begin{equation}
  \cnicemittingradius \, .
\end{equation}
In the shock frame, the frequency of the classical Colgate Noerdelinger radiation is
\begin{equation}
  \cnicshockedframecnfrequency
\end{equation}
The critical particle emission rate at which the plasma frequency exceeds the frequency in equation \ref{eq:cnic_shocked_frame_cn_frequency} is
\begin{equation}
  \cnicpprevanescentthreshold \, .
\end{equation}
At particle emission rates lower than equation \ref{eq:cnic_ppr_evanescent_threshold} (perdurable regime) the energy emitted by the classical Colgate Noerdelinger mechanism is given by
\begin{equation}
  \cniccnenergyperdurable \, .
\end{equation}
A fraction equal to the Thomson optical depth of these photons are Compton scattered, and their energy increases by a factor of $\gamma^2$, so the energy emitted in the form of Compton scattered photons is
\begin{equation}
  \cnicenergyperdurable \, .
\end{equation}
The fluence in the perdurable regime is given by
\begin{equation}
  \cnicfluenceperdurable
\end{equation}
From this equation we can calibrate the magnetic field
\begin{equation}
  \cnicemittingmagneticfieldperdurable
\end{equation}
which translates to neutron star surface magnetic field of
\begin{equation}
  \cnicneutronstarmagneticfieldperdurable \, .
\end{equation}

If the particle emission rates is higher than equation \ref{eq:cnic_ppr_evanescent_threshold} (evanescent regime) the energy emitted by the Colage Noerdelinger mechanism is given by
\begin{equation}
  \cniccnenergyevanescent \, .
\end{equation}
The calculation of the inverse Compton energy is the same as in the perdurable case, except photons can only travel a distance comparable to the plasma skin depth before they are absorbed instead of the entire shocked region. The emitted energy is therefore given by
\begin{equation}
  \cnicenergyevanescent \, .
\end{equation}
The fluence in the evanescent case is given by
\begin{equation}
  \cnicfluenceevanescent \, .
\end{equation}
Using this equation we can determine the magnetic field at the emitting radius
\begin{equation}
  \cnicemittingmagneticfieldperdurable
\end{equation}
which translates to neutron star surface magnetic field of
\begin{equation}
  \cnicneutronstarmagneticfieldevanescent \, .
\end{equation}
The reason equations \ref{eq:cnic_neutron_star_magnetic_field_perdurable} and \ref{eq:cnic_neutron_star_magnetic_field_evanescent} do not coincide at $\dot{N} = \dot{N}_{\rm ev}$ is that the transition is very sharp. when the radiation frequency drops below the plasma frequency, the thickness from which the radiation can be scattered drops suddenly from the rest thickness of the shell $r/\gamma$ to the skin depth $c/\omega_{\rm pl}$.

We proceed to calculate the energy required to drive such a shock. To do so, we have to determine whether the magnetic energy density dominates over the matter density or vice versa. In the perdurable regime, magnetic dominance occurs when
\begin{equation}
 \cnicpprmagneticdominanceperdurable \, . 
\end{equation}
Since this particle emission rate is higher than the threshold for the perdurable regime, then it is always in the magnetic dominated case.
In the evanescent case, the threshold is given by
\begin{equation}
  \cnicpprmagneticdominanceevanescent \, .
\end{equation}
The the power required to emit particles at the threshold for particle production rate in equation \ref{eq:cnic_ppr_magnetic_dominance_evanescent} is higher than the Eddington luminosity, and is therefore unphysical. The upshot is the upstream energy density is dominated by the magnetic field.

The shock energy in the perdurable case is
\begin{equation}
 \cnicexplicitcnenergyperdurable \, . 
\end{equation}
The shock energy in the evanescent case
\begin{equation}
  \cnicexplicitcnenergyevanescent \, .
\end{equation}
These energies exceed the energy budget of even magnetars, and therefore this mechanism cannot explain extragalactic FRBs. However, this mechanism might explain the FRB like emission from a galactic magnetar \citep{Andersen2020AMagnetar}. The isotropic equivalent luminosity of this event is lower by about four orders of magnitude compared with extragalactic FRBs. In addition, if we allow the particle emission rate to be as high as as that of the Crab pulsar, namely $10^{38} \, \rm s^{-1}$ \citep{Kennel1984ConfinementRemnant.}, we can shave off three more orders of magnitude and get to $10^{46} \, \rm erg$, which is close to the inferred energies of gamma ray flares \citep{Perna2011APhenomenology}.

\subsection{Counterparts}

\subsubsection{Panchromatic}

At smaller radii the shock wave was moving at a higher Lorentz factor. The highest Lorentz factor is attained at radii comparable to the light cylinder
\begin{equation}
  \cniclightcylinderfrequency \, .
\end{equation}
This frequency lies in the microwave range of the electromagnetic spectrum. In this range there are no time domain telescopes, and therefore this emission cannot be observed.

Another possibility of producing counterparts is by having photons experience two Compton scatterings. This boosts the frequency by a factor of $\gamma^2$, so that the frequency is
\begin{equation}
  \cnicdoublecomptonfrequency \, .
\end{equation}
This frequency is also in the microwave range, and so this transient cannot be observed.

\subsubsection{Incoherent}

Relativistic electrons moving in the presence of a magnetic field also produce synchrotron emission. Unlike the previous case, because of the high magnetic fields the cooling time of the electron is shorter than the dynamical time
\begin{equation}
  \cnicshockframesynchrotroncoolingtime \, .
\end{equation}
For this reason, we can assume fast cooling, i.e. that all the kinetic energy the electrons posses is emitted. This energy is given by
\begin{equation}
  \cnicsynchrotronfastcoolingenergy \, .
\end{equation}
The frequency at which this energy is emitted is
\begin{equation}
  \cnicobservedsynchrotronfrequency \, .
\end{equation}
This frequency corresponds to gamma rays, though higher particle emission rates the frequency can shift to the hard X ray range. Even for galactic events $d \approx 10 \, \rm kpc$, such an event corresponds to an instantaneous flux of $10^{-17} \, \rm \frac{erg}{cm^2 s}$, which is too dim for any X ray telescope.

We also note that synchrotron radio emission in this scenario is highly suppressed. This is because the observed cyclotron frequency
\begin{equation}
  \cnicobservedcyclotronfrequency
\end{equation}
is orders of magnitude higher than the radio range.

\section{Conclusions} \label{sec:conclusions}

In this work we present a new model for FRBs, based on insights from
observations of radio pulses from galactic neutron stars. In our
model, a sudden release of energy launches a plasma shell that
compresses the ambient magnetic field and emits coherent radio
signals. The key prerequisite, a
high bulk Lorenz factor, is motivated by recent Crab spectral Doppler
feature measurements\citep{Bij2021KinematicsPulses}, providing a
separation of the engine into a moving mirror that boosts ambient
magnetic fields into propagating radiation, and the separate
problem of acceleration of the mirror.  We do not address the latter problem,
leaving it for future scintillation observations.


Despite its simplicity, our model is able to explain a number of FRB features, namely the peak frequency, luminosity and drift to lower frequencies due to the deceleration of the shock. There are, however, a number features our simplistic model is unable to reproduce, in particular the difference in duration, bandwidth and peak flux between subsequent subpulses. Observations showed that different bursts can have opposite trends, and in some bursts trends can reverse throughout the burst \citep{Thornton2013ADistances, Hessels2019FRBStructure}. A simplistic model such as the one presented here is unable to reproduce such an inconsistent behaviour. We also note that in the burst from the galactic magnetar SGR 1935+2154 the peak frequency in the second subpulse is higher than the first, in contrast to all other known FRBs \citep{Andersen2020AMagnetar}. We note that the interval between the two subpulses is considerably longer than other FRBs (30 ms), which suggests that these are two subpulses might be two distinct explosions, with the second being more energetic. 

One non-generic result of our model is that the neutron stars that
produce fast radio bursts have relatively low magnetic fields, whereas
most detected galactic neutron stars have strong fields. This could
explain why offsets of localised FRBs (relative to their hosts'
centres) appear to be different from galactic neutron stars
\citep{Bhandari2022CharacterizingUniverse}. In addition, an FRB has
recently been detected in an old globular cluster
\citep{Kirsten2021ACluster}. Globular clusters are the hosts of
recycled millisecond pulsars \citep{Zhao2022AClusters}, whose
properties, namely rapid spin and low magnetic fields, are consistent
with our model.

That being said, we also show that a variant of our model can explain the FRB like emission from a galactic magnetar \citep{Andersen2020AMagnetar}. In this variant, coherent emission is first produced via the Colgate Noerdelinger mechanism at low frequency, and is subsequently inverse Compton scattered to the GHz range. We show that this variant requires too much energy to produce extragalactic FRBs but not for the galactic event, since the isotropic equivalent energy of the latter is four orders of magnitude lower than the former. In fact, we find that the required amount of energy is similar to the amount of energy released in gamma ray flares.



It has previously been pointed out that due to high latitude emission, FRB subpulses due to relativistic motion cannot drop too steeply \citep{Beniamini2020WhatMechanism}. This claim is based on a calculation that shows that if the shell stops emitting photons abruptly, because of differences in arrival time the flux measured by an observer will only decline as $t^{-2}$. However, we argue that a coherent emission can circumvent this limitation through interference. If we consider the electromagnetic field rather than the luminosity, then we can think of the effect of high latitude emission as a Green function that scales asymptotically as $t^{-1}$. For example, this means that a single pulse of emission $\delta \left(t\right)$ will produce a field that decays as $t^{-1}$, but two opposite pulses $\delta \left(t\right) - \delta \left(t + \Delta t\right)$ will tend to cancel each other and produce a field that decays as $t^{-2}$, and hence a flux that decays as $t^{-4}$. In a similar way it possible to obtain arbirarily steep observed fluxes

\section*{Acknowledgements}

AY is supported by the Natural Sciences and Engineering Research Council of Canada (NSERC), funding reference \#CITA 490888-16. This work made use of the sympy python package \citep{Meurer2017SymPy:Python}.

\section*{Data Availability}

This work does not make use of external databases.



\bibliographystyle{mnras}
\bibliography{mnras_template} 








\bsp	
\label{lastpage}
\end{document}